\documentclass[sigconf]{acmart}

\usepackage{booktabs}
\usepackage{graphicx}
\usepackage[inline]{enumitem} % for inline lists

\usepackage[T1]{fontenc}
\usepackage[scaled=0.85]{beramono}
\usepackage{listings}
\AtBeginDocument{%
  \providecommand\BibTeX{{%
    \normalfont B\kern-0.5em{\scshape i\kern-0.25em b}\kern-0.8em\TeX}}}

\renewcommand\footnotetextcopyrightpermission[1]{} % removes footnote with conference information in first column
\begin{document}


\title[TinyGenius: Intertwining NLP and Crowdsourced Microtasks]{TinyGenius: Intertwining Natural Language Processing with Microtask Crowdsourcing for Scholarly Knowledge Graph Creation}

\author{Allard Oelen}
\email{allard.oelen@tib.eu}
\orcid{0000-0001-9924-9153}
 \affiliation{
  \institution{TIB Leibniz Information Centre for Science and Technology}
   \city{Hannover}
   \country{Germany}
}
 
\author{Markus Stocker}
\email{markus.stocker@tib.eu}
\orcid{0000-0001-5492-3212}
 \affiliation{
  \institution{TIB Leibniz Information Centre for Science and Technology}
   \city{Hannover}
   \country{Germany}
}

\author{S\"oren Auer}
\email{soeren.auer@tib.eu}
\orcid{0000-0002-0698-2864}

 \affiliation{
  \institution{TIB Leibniz Information Centre for Science and Technology}
   \city{Hannover}
   \country{Germany}
}

\renewcommand{\shortauthors}{Oelen et al.}

\begin{abstract}
As the number of published scholarly articles grows steadily each year, new methods are needed to organize scholarly knowledge so that it can be more efficiently discovered and used. Natural Language Processing (NLP) techniques are able to autonomously process scholarly articles at scale and to create machine readable representations of the article content. However, autonomous NLP methods are by far not sufficiently accurate to create a high-quality knowledge graph. Yet quality is crucial for the graph to be useful in practice. We present TinyGenius, a methodology to validate NLP-extracted scholarly knowledge statements using microtasks performed with crowdsourcing. The scholarly context in which the crowd workers operate has multiple challenges. The explainability of the employed NLP methods is crucial to provide context in order to support the decision process of crowd workers. We employed TinyGenius to populate a paper-centric knowledge graph, using five distinct NLP methods. In the end, the resulting knowledge graph serves as a digital library for scholarly articles. 

%The approach is further evaluated by means of a user study where participants validate the generated NLP statements. 
\end{abstract}


\begin{teaserfigure}
  \includegraphics[width=\textwidth]{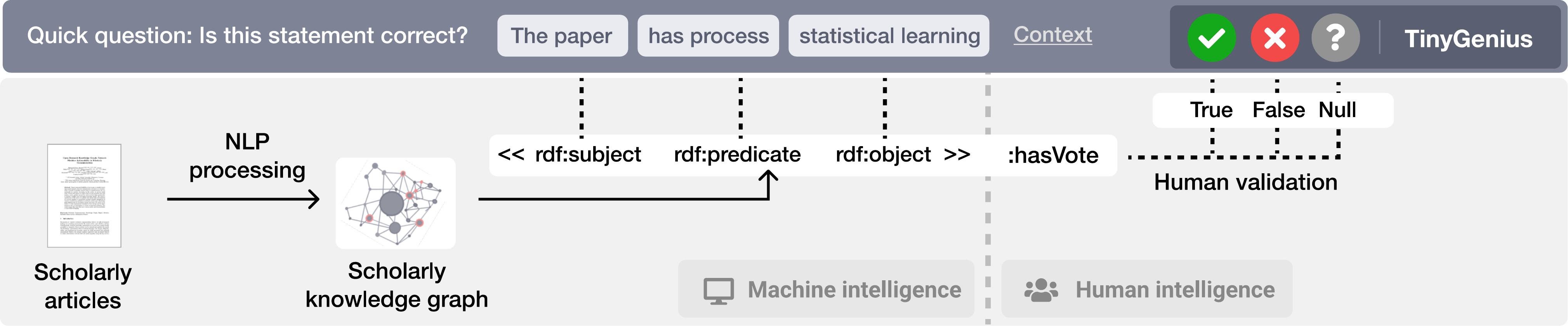}
  \caption{Graphical abstract. Workflow of the TinyGenius methodology. Scholarly articles are processed by NLP tools to form a scholarly knowledge graph (\textit{machine intelligence} part). Afterwards, the extracted statements are validated by humans by means of microtasks (\textit{human intelligence} part). User votes are stored as provenance data as part of the original statements.\\}
  \label{figure:teaser}
\end{teaserfigure}

\maketitle

\section{Introduction}

\begin{figure*}[t]
    \centering
    \includegraphics[width=1\textwidth]{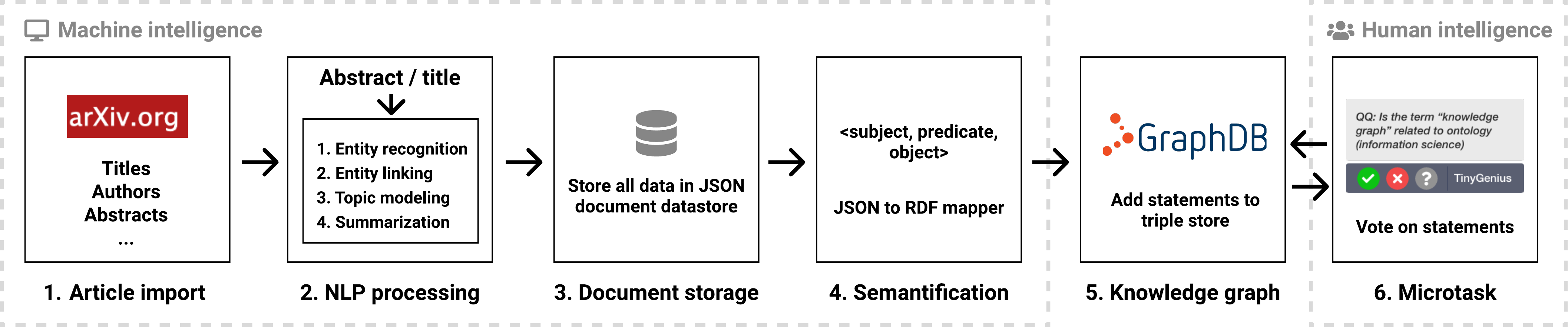}
    \caption{TinyGenius methodology intertwining human and machine intelligence to create a scholarly knowledge graph. ArXiv articles are imported, processed by a set of NLP tools, and the results are stored. From the results, a knowledge graph is generated. Afterwards, humans validate the knowledge graph by means of microtasks.}
    \label{figure:methodology}
\end{figure*}

Every year, the number of published scholarly articles grows~\cite{Jinha2010}, making it increasingly difficult to find and discover relevant literature. One of the key challenges is the ability of machines to interpret the knowledge presented within scholarly articles. Without \textit{machine actionable} scholarly knowledge, machines are severely limited in their utility to effectively organize this knowledge~\cite{Mons2009}. Knowledge graphs are a possible solution, as they enable knowledge to be represented in a machine readable manner. Knowledge graphs are foundational to scholarly digital libraries as they provide a means to efficiently discover and retrieve knowledge presented within research articles. 

In order to create a scholarly knowledge graph, structured knowledge has to be either extracted from the unstructured documents or produced directly upfront in the research workflow~\cite{Stocker_2018}. We distinguish between two different strategies to support the extraction process. Firstly, there is manual structured knowledge extraction with human labor. This will most likely result in high-quality data, however this approach does not scale well. Secondly, there is automatic extraction using machine learning techniques. Specifically, Natural Language Processing (NLP) is able to interpret natural language and transform unstructured content into a structured, machine readable representation. However, NLP tools are not sufficiently accurate to generate a high-quality knowledge graph, in particular, due to the complexity of the conveyed information, the required context-awareness or the varying levels of semantic granularity. In this work, we propose a hybrid method where we combine human and machine intelligence via microtasks to create a structured scholarly knowledge graph. 

We present \textit{TinyGenius}, a methodology to create a scholarly knowledge graph leveraging intertwined human \textit{and} machine intelligence. Firstly, NLP tools are used to autonomously process scholarly articles. Secondly, the NLP results are transformed into a paper-centric scholarly knowledge graph. Finally, the statements are presented to humans in the form of microtasks. Humans can vote to determine the correctness of the statements. Based on the votes, an aggregated score is computed to indicate the correctness of a statement. TinyGenius is specifically designed to be integrated into the Open Research Knowledge Graph (ORKG)~\cite{Jaradeh2019a}. The ORKG leverages a crowdsourcing approach to curate a scholarly knowledge graph~\cite{10.1145/3397481.3450685}.

\section{Related Work}
Large complex tasks can be decomposed into a set of smaller, independent microtasks~\cite{10.1145/2642918.2647349}. These microtasks are context-free, are more manageable, and are generating higher quality results~\cite{10.1145/2702123.2702146}. While microtasks can be beneficial on an individual level, such as microwork~\cite{10.1145/2559206.2581181}, they are commonly performed in a crowdsourced setting by unskilled users~\cite{10.1007/978-3-642-35176-1_33}. In a crowdsourced setting, a large task, too big in scope for a single person, can be completed collaboratively. Microtask crowdsourcing has been successfully employed for various tasks, for example, writing software programs~\cite{10.1145/2642918.2647349}, validating user interfaces~\cite{10.1145/2470654.2470684}, labeling machine learning datasets~\cite{10.1145/3025453.3026044}, ontology alignment~\cite{10.1007/978-3-642-35176-1_33}, and knowledge graph population~\cite{10.1093/bioinformatics/btt333}.

\begin{table*}[t]
\centering
\caption{List of employed NLP tools and their corresponding task and scope. The question template shows how the microtask is presented to the user. }
\label{table:nlp-tools}
\resizebox{0.9\textwidth}{!}{%
\begin{tabular}{@{}llll@{}}
\toprule
\textbf{Tool name} & \textbf{NLP task} & \textbf{Scope} & \textbf{Question template} \\ \midrule
\textit{CSO classifier} & Topic Modeling & Domain-specific & Is this paper related to the topic \{topic\}? \\
\textit{Ambiverse NLU} & Entity Linking & Generic & Is the term \{entity\} related to \{wikidata concept\}? \\
\textit{Abstract annotator} & Named Entity   Recognition & Domain-specific & Is this statement correct? This paper \{type\} \{entity\} \\
\textit{Title parser} & Named Entity   Recognition & Domain-specific & Is \{entity\} a \{type\} presented in this paper? \\
\textit{Summarizer} & Text Summarization & Generic & Does this summarize the paper correctly? \\ \bottomrule
\end{tabular}%
}
\end{table*}

Machine learning tools are able to process data at scale without the need for human assistance. Therefore, such tools are especially suitable to handle large quantities of data, such as scholarly article corpora. The Natural Language Processing (NLP) domain focuses specifically on understanding natural language for machines~\cite{https://doi.org/10.1002/aris.1440370103}. In our methodology, we employ a set of five NLP tools to process scholarly article text. These tools perform four different NLP tasks, which we will now discuss in more detail. First, \textit{Named Entity Recognition} (NER) is a task to identify entities within text belonging to a predefined class~\cite{9039685}. Second, \textit{Entity Linking} is the task of linking entities to their respective entry in a knowledge base~\cite{6823700}. Third, \textit{Topic Modeling} is the task to identify and distinguish between common topics occurring in natural text~\cite{Alghamdi2015}. Finally, \textit{Text Summarization} is the task of compressing text into a shorter form, while preserving the key points from the original text~\cite{tas2007survey}.

\section{Architecture and User Interface}
We now discuss the TinyGenius methodology. First, we focus on the technical infrastructure that is responsible for data storage and processing. Afterwards, we explain the user interface in more detail.  The data model relies on triple statements using the W3C Resource Description Framework (RDF)~\cite{lassila1998resource}. By using a standardized data representation model, the data interchange between machines is facilitated. RDF data can be queried using the SPARQL language~\cite{prudhommeaux2008sparql}.

\begin{figure*}[t]
    \centering
    \includegraphics[width=0.82\textwidth]{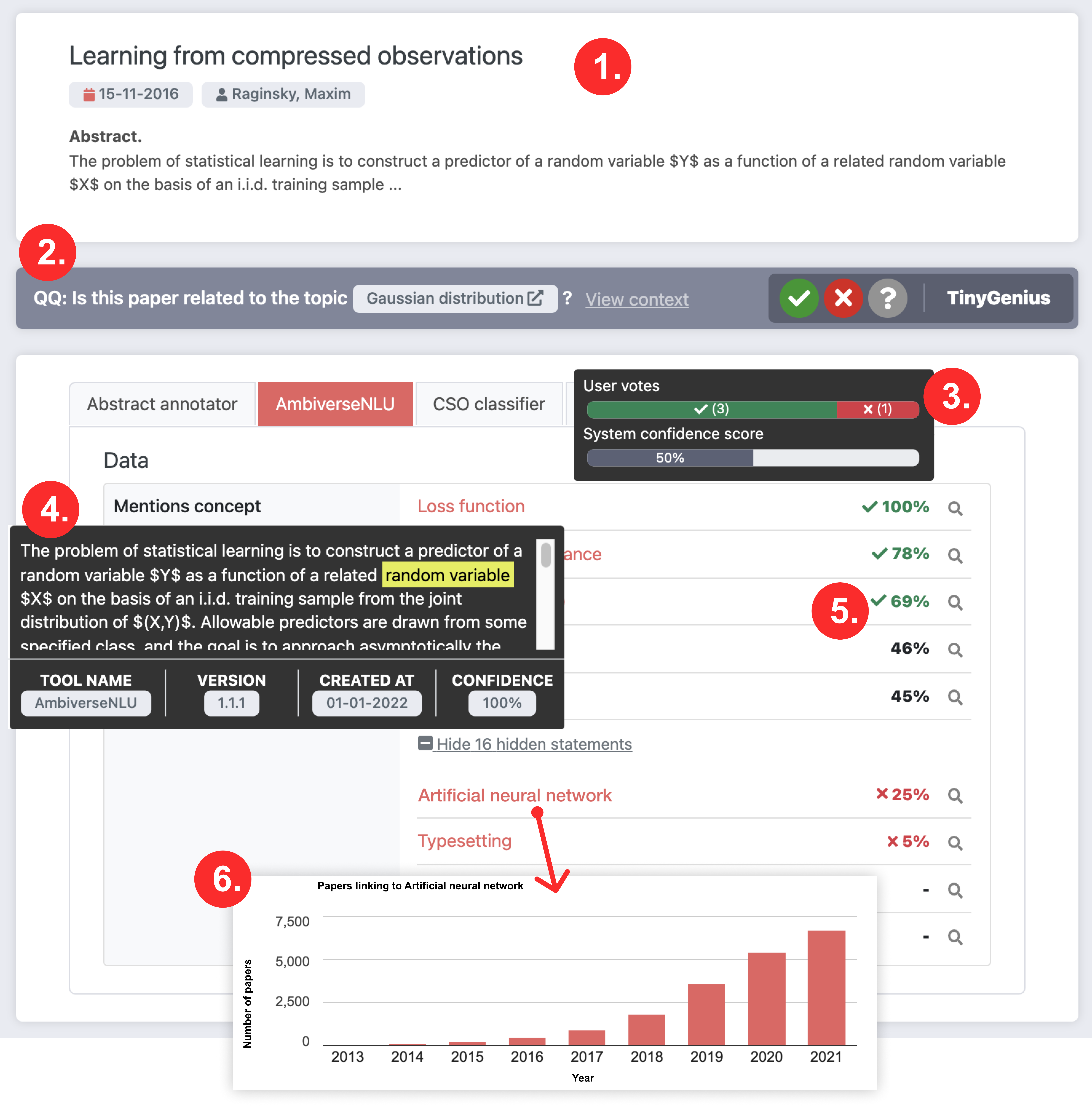}
    \caption{View paper page, showing the integrated voting widget and NLP statements. Node 1 displays the metadata related to the selected paper. Node 2 shows the voting widget. Node 3 is the score tooltip. Node 4 shows a tooltip that displays the context and provenance data related to a single statement. Node 5 lists the NLP-generated statements grouped by the tool. Finally, node 6 shows the use of a resource grouped by year, which is displayed when clicking on a resource.}
    \label{figure:view-paper-page}
\end{figure*}
\subsection{Technical Infrastructure}
One of the key benefits of using NLP tools to process data is the ability to perform this analysis at scale. Therefore, the infrastructure is designed to handle large quantities of data while still performing well. We outline the methodology depicted in Figure~\ref{figure:methodology}:
\begin{enumerate}
    \item In the first step, the complete metadata corpus from the open-access repository service arXiv\footnote{\url{https://arxiv.org/}} is imported. This includes article titles and abstracts. To reduce the required computational resources and ensure a consistent level of semantic granularity, only paper titles and abstracts are processed by NLP tools (i.e., the full-text is excluded). 
    \item Afterwards, the papers are processed by different NLP tools, which are listed in Table~\ref{table:nlp-tools}.
    \item In the third step, the output of the paper import process and the resulting data from the NLP tools are stored in a document-based JSON data store. Notably, the NLP results are stored in their native data model and are not transformed to make them suitable for knowledge graph ingestion. 
    \item The semantic transformation process takes place in the fourth step, i.e. semantification. This step converts the native NLP data models to a triple format, as required by the RDF data model. 
    \item In the fifth step, the data is ingested in a triple store. We adopted an RDF*~\cite{Hartig1141963} provenance data model. Therefore, a GraphDB\footnote{\url{https://graphdb.ontotext.com/}} triple store is used, which supports RDF* natively. To increase machine-actionability, existing ontologies concepts are used when possible. 
\end{enumerate}

\subsection{User Interface}
The user interface consists of two main components: the view paper page and the voting widget. Figure~\ref{figure:view-paper-page} shows a screenshot of the view paper page. It shows how a single paper is displayed when integrated within the ORKG. All data displayed on the page is coming from the TinyGenius knowledge graph and is fetched using SPARQL. The voting widget is the key interface component and integrates the microtasks to perform the NLP validation. It is displayed in Figure~\ref{figure:view-paper-page} node 2. Each NLP tool has a different question template, as listed in Table~\ref{table:nlp-tools}. This question template is used to display the microtask in the widget. The widget itself displays the context required to make an informed decision about the correctness of the statement. In most cases, the context displays an excerpt of the abstract and highlights the words used by the NLP tool to extract the data. Finally, users are able to vote about the correctness. A vote can either be correct, incorrect, or unknown. The next statement is automatically displayed after voting. Statements are selected in random order and statements are only displayed once to a specific user. By default, statements with a score below a certain threshold (40\%) are hidden within the user interface.

\section{Data Evaluation}
We conduct a data evaluation to gather general statistics about our approach and to assess the technical performance of the system. To this end, we imported the arXiv corpus and processed a subset with selected NLP tools. All articles classified as ``Machine Learning'' by arXiv are processed. This results in a total amount of $95,376$ processed articles, which is approximately 5\% of the complete arXiv corpus. We consider this a sizable amount to estimate statistics such as processing time per article, number of extracted statements per article, and to determine the performance of the setup. We chose the machine learning field because several NLP tools are trained specifically on machine learning abstracts. The processing time in seconds per NLP tool is listed in Table~\ref{table:data-evaluation}. In addition to the total number of triples, an approximation of the number of metadata and provenance triples is listed. The tools ran on a machine with 40 CPU cores and no dedicated GPUs. As the summarizer tool requires GPUs to run efficiently, we did not apply this tool to the dataset. 

\begin{table}[t]
\caption{Overview of the data evaluation statistics.}
\label{table:data-evaluation}
\resizebox{0.75\columnwidth}{!}{
\begin{tabular}{@{}l|r@{}}
\toprule
\textbf{Description} & \multicolumn{1}{l}{\textbf{Measure}} \\ \midrule
\textit{General statistics} & \multicolumn{1}{r}{\textit{Number}} \\
Processed articles & 95,376 \\
Triples metadata & 1,521,492 \\
Triples provenance & 47,595,706 \\
Triples total & 65,608,902 \\
Average number of   triples per article & 688 \\
\midrule
\textit{Processing time} & \multicolumn{1}{r}{\textit{Seconds}} \\
\midrule
CSO classifier & 27,803 \\
Ambiverse NLU & 137,060 \\
Abstract annotator & 62,056 \\
Title parser & 87 \\
Summarizer & N/A \\ \bottomrule
\end{tabular}}
\end{table}

\section{Discussion and Conclusion}
We presented TinyGenius, a methodology to validate NLP statements using microtasks. The method combines machine and human intelligence resulting in a synergy that utilizes the strengths of both approaches. Firstly, a set of NLP tools is applied to a corpus of paper abstracts. Afterwards, the resulting data is ingested in a scholarly knowledge graph. Finally, the data is presented to users in the form of microtasks. By utilizing microtasks, the data is validated using human intelligence. We envision our approach to be integrated within the ORKG, presenting the microtasks to ORKG users (generally researchers). The ORKG already leverages crowdsourcing to curate knowledge and by introducing microtasks we lower the barrier to become a \textit{content contributors} for users that are normally merely \textit{content consumers}. 

The preliminary data evaluation results indicate that the presented method is promising and the proposed setup and infrastructure are suitable for the task. When the methodology is deployed in a real-life setting, the knowledge graph quality can be substantially improved. Over time, more visitors will vote on the presented statements, increasing the overall data accuracy. The user votes are stored as provenance data on the statement level, providing the opportunity for downstream applications to decide how to incorporate the validation data. Incorrect data can simply be filtered out, but it is also possible to perform more complex analysis on the validation data. 

Within this work, we laid the foundation for a comprehensive scholarly knowledge infrastructure. A more in-depth evaluation is part of future work, including an analysis of the system performance, a user evaluation in a controlled environment, and an evaluation when deployed in scholarly knowledge platform. Especially, the latter evaluation will give insights on how the approach provides benefits to researchers and how it can be used to form a digital library for scholarly articles. We will specifically focus on creating tools and interfaces to support scholarly knowledge discovery, for example via dynamic faceted search tools. Additionally, we will focus on trend analysis, in the form of scientometrics. 

The approach has been evaluated with machine learning articles from the arXiv corpus. Some of the selected NLP tools are domain models, specifically trained on Computer Science. However, our approach is not limited to this domain. By design, the system is modular and can be generalized to support other domains and NLP tools. Future work will focus on importing the complete arXiv corpus, which increases the number of triples approximately tenfold.

We deem this work to be one of the first, which truly combines human and machine intelligence for knowledge graph creation and curation. This combination needs much more attention, since there are many use cases, where machine intelligence alone can (due to the missing training data) not produce useful results.

\begin{acks}
This work was co-funded by the European Research Council for the project ScienceGRAPH (Grant agreement ID: 819536) and the TIB Leibniz Information Centre for Science and Technology. We would like to thank Mohamad Yaser Jaradeh and Jennifer D'Souza for their contributions to this work.
\end{acks}
\bibliographystyle{ACM-Reference-Format}
\bibliography{refs}

\end{document}